\documentclass[10pt,aps,prl,twocolumn,floatfix]{revtex4-2}
                        \usepackage{graphicx}
                        \usepackage{dcolumn}
                        
\begin{document}

                        \def\be{\begin{equation}}
                        \def\ee{\end{equation}}
                        \def\ba{\begin{eqnarray}}
                        \def\ea{\end{eqnarray}}
                        \def\bas{\begin{eqnarray*}}
                        \def\eas{\end{eqnarray*}}


\title{Piecewise Parabolic Kinetic Energies and Classical Time Crystals}

\author{Stavros Theodorakis and Athina Hadjizorzi}
                        \affiliation{Physics Department, University of Cyprus,
P.O. Box 20537, Nicosia 1678, Cyprus}
                        \email{stavrost@ucy.ac.cy}
\date{\today}

\begin{abstract}
We present explicit, exact analytic solutions for classical time crystals embedded in a parabolic potential for the position. While Lagrangians with a quartic kinetic energy yield complex trajectories governed by inscrutable elliptic integrals, we demonstrate that unconventional piecewise parabolic kinetic energies —minimised at nonzero values of the speed— admit simple, transparent periodic solutions. In these systems, the speed acquires definite periodic discontinuities but remains always nonzero. The requirement that the action be the smallest available determines the branch that the solution will follow, ensuring in addition that it is stable there and rendering the Hamiltonian single-valued as a function of momentum. Furthermore, we introduce a powerful piecewise parabolic emulation framework to approximate highly non-linear kinetic energies, demonstrating its utility on both quartic time crystals and the relativistic harmonic oscillator. We show that finding exact solutions to these emulated models provides accurate, physically transparent simulacra of complicated problems, offering a robust alternative to standard numerical approximations for non-quadratic kinetic energies. 
\end{abstract}

\maketitle

\vskip 0.3cm
\vskip 0.3cm

{\bf I. Introduction}

In 2012 Shapere and Wilczek\cite{Wilczek} investigated classical Lagrangians in which the kinetic energy was minimised at nonzero speeds, displaying thus motion in the lowest energy state. This tendency of the kinetic energy was counteracted though by the tendency of the potential energy to have a static minimum at a particular position. These two contradictory tendencies can be accommodated simultaneously by zigzag "brick wall" solutions, in which the speed suddenly reverses direction, this reversal conserving nonetheless the energy. For initial speeds very close to the minimum of the kinetic part of the Hamiltonian the orbits ricochet about this minimum with very small amplitude and very small period. In that limit we have an autonomous, time periodic solution of the equations of motion that is simultaneously a local minimum of the energy. The systems that display such behaviour were called time crystals.  Time crystalline behaviour, with a spontaneously broken time translation symmetry, appears also in Hamiltonian systems with conserved quantities, where the numerical values of the conserved charges are constrained\cite{Niemi}.

The Lagrangian examined in Ref. \cite{Wilczek} had a quartic kinetic energy and a general potential energy. However, the main features of these zigzag solutions, the so called Sisyphean dynamics\cite{Sisyphus}, appear in many other Lagrangians with unconventional kinetic energies\cite{symmetry}. In particular,  they can appear in Lagrangians with piecewise parabolic functions of the speed as their kinetic energies, leading thus to simple analytic solutions.

In this paper we present a full analytic study of two such cases, one with a piecewise parabolic Lagrangian and another with a piecewise parabolic Hamiltonian, elucidating fully the zigzag nature of the solutions and their periodicities. We also examine a Lagrangian with quartic kinetic energy. We introduce a piecewise parabolic emulation of the quartic kinetic energy that leads to simple equations of motion, the solutions of which approximate very adequately the complicated exact solutions of the quartic case. This method of emulation can be used for any complicated kinetic energies or potentials, leading thus to simple approximations for the exact solutions of complicated problems. In other words, instead of trying to find approximate solutions to the original problem, we find exact solutions to the emulated problem. In this paper, all the work is done, without loss of generality, for a simple harmonic oscillator potential for the position.

Let us then ask first ourselves what kind of kinetic energies would give zigzag solutions that would vacillate very quickly around the position $x(t)=0$, the minimum of the simple harmonic potential. In such solutions the speed $v(t)$ should be able to change sign instantaneously at a particular instant. This means that there would be a finite discontinuity in the speed at that point and consequently the acceleration there would be infinite. Let us consider then the Lagrangian $L=T(v(t))-x(t)^2/2$, where $T(v(t))$ is the kinetic energy. Its equation of motion takes the form $dv/dt=-x(t)/(\partial^{2}T/\partial v^{2})$. Thus the acceleration will be infinite whenever the second partial derivative $\partial^{2}T/\partial v^{2}$ of the kinetic energy with respect to the speed becomes zero. These points will determine the edges of the vacillation of the solutions, since the speed will suddenly change sign there and $|x(t)|$ will stop increasing and start decreasing.

Furthermore, let us examine the corresponding Hamiltonian: $H=h_{K}+V$, where $V=x(t)^2/2$ is the potential energy and where $h_{K}=v(\partial T/\partial v)-T$ is the kinetic part of the Hamiltonian. This kinetic part $h_{K}$ of $H$ has an extremum when $\partial^{2}T/\partial v^{2}=0$, i.e. when the acceleration is infinite. As an example, the choice $T(v)=v^4/4-v^2/2$ of Ref. \cite{Wilczek} has infinite acceleration at $v=\pm 1/\sqrt{3}$; at these values of the speed the kinetic part of the corresponding Hamiltonian is indeed minimised.

Conventional quadratic kinetic energies will not display time crystalline behaviour, since for them $\partial^{2}T/\partial v^{2}$ is never zero. We shall see, however, in the next section, that the zigzag behaviour can arise in piecewise quadratic kinetic energies.

{\bf II. A piecewise parabolic kinetic energy}

\begin{figure}[t]
\vskip 0.3cm
 \includegraphics[width=0.47\textwidth]{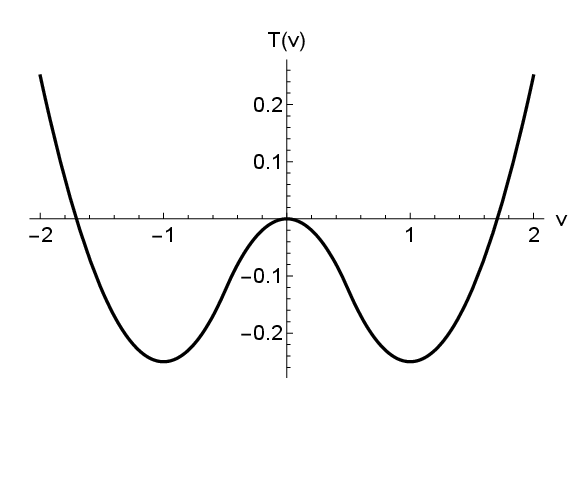}
\caption{\label{fig1}The piecewise parabolic kinetic energy $T(v)$
of Eq.~(\ref{parabolickinetic}) as a function of the speed.}
\end{figure}

Let us examine the following kinetic energy $T(v)$, seen in Figure~\ref{fig1}, that displays spontaneous symmetry breaking:

\ba
\label{parabolickinetic}
T(v)&=&(v-1)^2/2-1/4\,\,\,\,\,\,\,\,\,if\,\,\,\,v\geq1/2\nonumber\\
&=&-v^2/2\,\,\,\,\,\,\,\,\,\,if\,\,\,\,\,1/2\geq v\geq -1/2\nonumber\\
&=&(v+1)^2/2-1/4\,\,\,\,\,if\,\,\,\,\,-1/2\geq v
\ea

This piecewise parabolic kinetic energy and its slope are continuous for every value of the speed. We note that a piecewise quartic kinetic energy has already been studied\cite{Chi}.                       
                        
\begin{figure}[t]
\vskip 0.3cm
                        \includegraphics[width=0.47\textwidth]{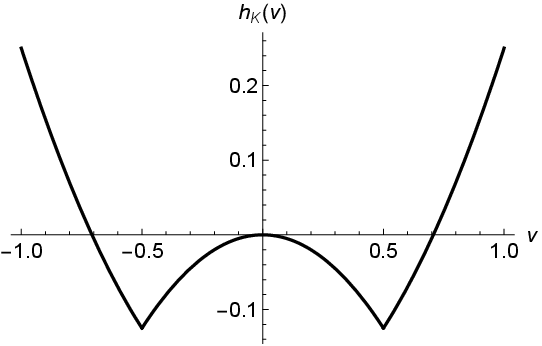}
                        \caption{\label{fig2}The piecewise parabolic kinetic part of the Hamiltonian $h_{K}(v)$ of Eq.~(\ref{parabolickinetichamiltonian}) as a function of the speed.}
                        \end{figure}

                        The corresponding Hamiltonian is $h_{K}(v)+x(t)^2/2$, where the kinetic part $h_{K}$ of the Hamiltonian, shown in Figure~\ref{fig2}, is:
                        
                   \ba
\label{parabolickinetichamiltonian}
h_{K}(v)&=&(2v^2-1)/4\,\,\,\,\,\,\,\,\,if\,\,\,\,v\geq1/2\nonumber\\
&=&-v^2/2\,\,\,\,\,\,\,\,\,\,if\,\,\,\,\,1/2\geq v\geq -1/2\nonumber\\
&=&(2v^2-1)/4\,\,\,\,\,if\,\,\,\,\,-1/2\geq v
\ea

										This kinetic part of the Hamiltonian is minimised at the cusps $v=\pm 1/2$.  Note that the energy is multi-valued in terms of the canonical phase space variables and the symmetry breaking ground states are all located at the branching point singularities\cite{Chinese}.	We note also that the kinetic energy of Eq.~(\ref{parabolickinetic}) can be written in the form:
										
\ba
\label{heaviside}
&&(v-\frac{1}{2})^2\theta(v-\frac{1}{2})-(v+\frac{1}{2})^2\theta(v+\frac{1}{2})\nonumber\\
&&+v^2/2+v+\frac{1}{4},
\ea

where $\theta(x)$ is the Heaviside $\theta$ function, with $\theta(x)=1$ if $x>0$, $\theta(0)=1/2$ and $\theta(x)=0$ if $x<0$. The second derivative of this function of $v$ is zero at $v=\pm 1/2$, therefore we expect discontinuities in the speed there. The corresponding kinetic part of the Hamiltonian is
				
\be
\label{heavisidehamiltonian}
h_{K}(v)=\frac{2v^2-1}{4}+(v^2-\frac{1}{4})(\theta(v-\frac{1}{2})-\theta(v+\frac{1}{2})).
\ee

We note that $h^{\prime}_{K}(\pm 1/2)=0$, while $h^{\prime\prime}_{K}(\pm 1/2)$ is infinitely positive. Thus the kinetic part of the Hamiltonian is indeed minimised at $v=\pm 1/2$.

If we multiply the equation of motion with $v(t)$ and integrate with respect to the time, we shall see that the Hamiltonian $H=h_{K}+x(t)^2/2$ is a constant, the energy. We adopt, without loss of generality, the initial conditions $x(0)=0$ and $v(0)=1/2+\epsilon$, where $\epsilon>0$. Thus the initial position is the minimum of the potential energy, while the initial speed can be very close to a minimum of the kinetic part of the energy. Then the total energy is equal to $E=(\epsilon^{2}+\epsilon)/2-1/8$. Energy conservation implies thus that at the positions $\pm\sqrt{\epsilon^2+\epsilon}=\pm A$ the square of the speed is 1/4.

The equation of motion is $\ddot{x}(t)=\pm x(t)$, where the upper sign applies to the region $1/2>v>-1/2$, while the lower sign applies to the regions $v>1/2$ and $-1/2>v$. Since the particle began its motion at $t=0$ with speed $1/2+\epsilon$, the corresponding equation of motion (in the region $v>1/2$) is $\ddot{x}+x=0$. The solution is

\be
\label{x1}
x_{1}(t)=(\frac{1}{2}+\epsilon)\sin t
\ee
if $0\leq t\leq t_{1}$.

We define the time $t_{1}$ through the equations
\be
\label{sint1}
\sin t_{1}=\sqrt{\epsilon^2+\epsilon}/(1/2+\epsilon)
\ee
and
\be
\label{cost1}
\cos t_{1}=(1/2)/(1/2+\epsilon).
\ee
Then $x_{1}(t_{1})=-\ddot{x}_{1}(t_{1})=\sqrt{\epsilon^2+\epsilon}$ and $\dot{x}_{1}(t_{1})=1/2$.

Hence, at time $t_{1}$ the solution has reached the boundary of the region $v>1/2$, at the position $\sqrt{\epsilon^2+\epsilon}=A$, with a negative acceleration $-A$. As it crosses this boundary, we expect the position to maintain its value $A$, with the corresponding speed being $\pm 1/2$ there. The speed, however, must have a discontinuity, as we explained before. Indeed, at the very instant that it crosses into the region $1/2>v>-1/2$, the applicable equation of motion will be $\ddot{x}-x=0$ and hence the slope of the speed will change to $+A$. The only way this increasing speed will remain in the intermediate region $1/2>v>-1/2$ is if it jumps to the other value of the speed that corresponds to the value $x=A$, i.e. the value $-1/2$. There is of course the possibility that it jumps to the region $-1/2>v$. In both cases, however, the speed will jump at time $t_{1}$ to the value $-1/2$. Thus a different branch of the motion commences there. This branch could proceed to values of the speed lower than $-1/2$ or values of the speed greater than $-1/2$.

In fact, whenever the solution reaches the edge of the vacillation, with $|x(t)|=A$ and $|v|=1/2$, energy conservation forces the speed to jump to $1/2$ or $-1/2$. But as it proceeds further in time, it has two options: to move above or below this newly acquired value of the speed. Thus two branches appear as options after each jump.  This seems to imply that one cannot predict the classical motion of the system if the initial position and speed is given, since at particular instants it can switch from one branch to the other. In fact, one could visualize the classical motion as a succession of zigzags which happen in an unpredictable manner\cite{Henneaux}.

We can, however, determine the branch the system will finally follow if we impose the requirement that the branch chosen have the smallest available value for the action in the neigbourhood of the branching point\cite{admissibility}. This principle of the smallest available action must hold for each individual infinitesimal segment of the path taken. If we calculate the action for the time interval $(t_{0},t_{0}+\Delta t)$, where $\Delta t$ is small and positive, the preferred branch will be the one corresponding to the smaller action. In fact, the requirement that the action be minimized over an infinitesimal path has been used to derive the Euler Lagrange equations and the conservation of energy using simple derivatives\cite{Hanc}.

We shall apply this requirement to the branches $|v|>1/2$ first. For given initial conditions $x(t_{0})=x_{0}$ and $\dot x(t_{0})=v_{0}$ we shall have

\be
\label{genx}
x(t)=x_{0}\cos(t-t_{0})+v_{0}\sin(t-t_{0}).
\ee

For the branch $\beta_{+}$ ($v>1/2$) the action for the interval $(t_{0},t_{0}+\Delta t)$ is

\be
\label{Ssyn}
S_{+}\approx\frac{1}{4}(1-4v_{0}+2v_{0}^2-2x_{0}^2)\Delta t+\frac{1}{2}x_{0}(1-2v_{0})\Delta t^2.
\ee

For the branch $\beta_{-}$ ($v<-1/2$) the action for the interval $(t_{0},t_{0}+\Delta t)$ is

\be
\label{Splin}
S_{-}\approx\frac{1}{4}(1+4v_{0}+2v_{0}^2-2x_{0}^2)\Delta t-\frac{1}{2}x_{0}(1+2v_{0})\Delta t^2.
\ee

Finally, for the branch $\beta_{0}$ ($1/2>v>-1/2$), with the same initial conditions and hence with 

\be
x(t)=x_{0}\cosh(t-t_{0})+v_{0}\sinh(t-t_{0}). 
\ee

the action for the interval $(t_{0},t_{0}+\Delta t)$ is

\be
\label{S0}
S_{0}\approx\frac{1}{2}(-v_{0}^2-x_{0}^2)\Delta t-x_{0}v_{0}\Delta t^2.
\ee

\begin{figure}[t]
\vskip 0.3cm
                        \includegraphics[width=0.47\textwidth]{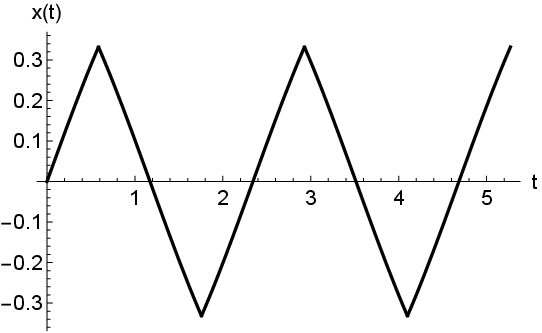}
                        \caption{\label{fig3} A periodic solution for the position $x(t)$ for $\epsilon=0.1$, arising from the interchange of the speed between the regions $v>1/2$ and $v<-1/2$.}
                        \end{figure}

\begin{figure}[t]
\vskip 0.3cm
                        \includegraphics[width=0.47\textwidth]{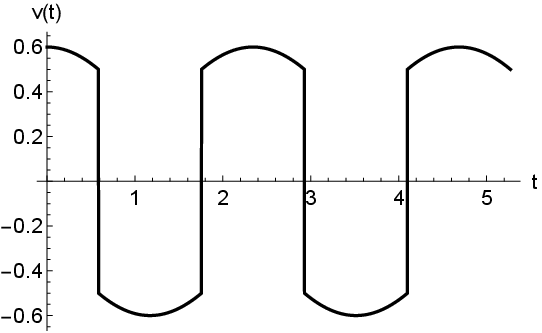}
                        \caption{\label{fig4} A periodic solution for the speed $v(t)$ for $\epsilon=0.1$, arising from the interchange of the speed between the regions $v>1/2$ and $v<-1/2$.}
                        \end{figure}

Let us examine then the system at time $t_{1}$. The speed has to jump to the value $-1/2$, so the initial conditions are $x(t_{1})=A$ and $\dot x(t_{1})=-1/2$. This is the boundary of regions $\beta_{0}$ and $\beta_{-}$. Which one will the system prefer? We see that for these initial conditions $S_{0}-S_{-}\approx A\Delta t^{2}/2$. Hence the system will choose the branch $\beta_{-}$, which has the lower action. The solution takes then the sinusoidal form

\be
\label{x2}
x_{2}(t)=-(\frac{1}{2}+\epsilon)\sin(t-2t_{1})
\ee

for $t_{1}\leq t\leq 3t_{1}$. We note that $x_{2}(3t_{1})=-A$ and $\dot{x}_{2}(3t_{1})=-1/2$. Thus at time $3t_{1}$ the speed has returned to the value $-1/2$ and it will instantaneously jump to $1/2$. Hence the initial conditions for this next part of the motion will be  $x(3t_{1})=-A$ and $\dot{x}(3t_{1})=1/2$. Correspondingly, the difference in the actions of the two possible branches, which have $v=1/2$ as their boundary, is $S_{0}-S_{+}\approx A\Delta t^{2}/2$. This means that the preferred branch is the $\beta_{+}$. At the moment $3t_{1}$ the solution proceeds then into the region $\beta_{+}$. Since the system is in the sinusoidal region, $v>1/2$, this next piece of the resulting periodic solution, portrayed for the position in Figure~\ref{fig3} and for the speed in Figure~\ref{fig4}, will be
	
\be
\label{x3}
x_{3}(t)=(\frac{1}{2}+\epsilon)\sin(t-4t_{1})
\ee
if $3t_{1}\leq t\leq 5t_{1}$. It then jumps back to the speed $-1/2$, and so forth.

In general,

\be
\label{xn}
x_{n}(t)=(-1)^{n+1}(\frac{1}{2}+\epsilon)\sin(t-2(n-1)t_{1}).
\ee
if $(2n-3)t_{1}\leq t\leq (2n-1)t_{1}$.

In other words, every time the speed acquires the value $\pm 1/2$ it immediately jumps to the value $\mp 1/2$. We note in Figure~\ref{fig3} and  Figure~\ref{fig4} that the position and the speed have the same sign just before each jump. The period of the oscillation is $4t_{1}$, while the amplitude of the full oscillation is $2A$. For small $\epsilon$ the amplitude is approximately $2\sqrt{\epsilon}$, while the semiperiod is $4\sqrt{\epsilon}$, both of them very small. The mean speed for such small $\epsilon$ is, of course, 1/2.

As we have seen then, the principle of the smallest available action ensures that the $|v|<1/2$ region, where the solutions are hyperbolic functions, is forbidden.

Furthermore, we note that the $|v|<1/2$ branch is unstable. Indeed, if $S$ is the action for a solution $x(t)$ of the equations of motion in this region, we can perturb that solution to $x(t)+\epsilon\eta(t)$ and obtain the action $S_{0}+\epsilon\,\,\delta S+\epsilon^{2}\delta^{2}S/2$. The second variation is

\ba
&&\delta^{2}S=\frac{d^2}{d\epsilon^2}S(x+\epsilon\eta)|_{\epsilon=0}\nonumber\\
&&=\frac{d^2}{d\epsilon^2}\int\,dt\left(-(x+\epsilon\eta)^{2}/2-(\dot x+\epsilon\dot \eta)^2/2\right)|_{\epsilon=0}\nonumber\\
&&=-\int\,dt\left(\eta^2+\dot \eta^2\right).
\ea

The solutions for this $\eta$ are hyperbolic functions. Therefore the branch $|v|<1/2$ is unstable.

In contrast, for the sinusoidal branch the second variation is

\be
\delta^{2}S=\int\,dt\left(-\eta^2+\dot \eta^2\right),
\ee

which corresponds to plain oscillations. Therefore that branch is stable.

In fact, even if we had started with a hyperbolic solution at $t=0$, at the first jump of the speed the solution would become sinusoidal. Indeed, if the initial conditions were $x(0)=0$ and $v(0)=1/2-\epsilon$, with $\epsilon>0$, the solution for $0\leq t\leq\tau_{1}$ would be $x(t)=(1/2-\epsilon)\sinh(t)$, where $\tau_{1}=\cosh^{-1}(1/(1-2\epsilon))$. The position and the speed at time $\tau_{1}$ would be $x(\tau_{1})=\sqrt{\epsilon-\epsilon^2}=A_{h}$ and $\dot x(\tau_{1})=1/2$. At that point the speed would have to jump to the value $-1/2$. At that fork of the road the solution would have the option of going into the hyperbolic $\beta_{0}$ ($|v|<1/2$) region or into the sinusoidal $\beta_{-}$ ($-1/2>v$) region. However, with the initial conditions $x_{0}=A_{h}$ and $v_{0}=-1/2$ we have $S_{0}-S_{-}\approx A_{h}\Delta t^{2}/2$. Hence the principle of the smallest available action will force the system to advance into the sinusoidal $\beta_{-}$ region. In other words, even with a hyperbolic beginning, we would end up almost immediately on the sinusoidal branch.

{\bf III. The piecewise parabolic Hamiltonian as a function of momentum}

The kinetic part of the piecewise parabolic Hamiltonian of Eq.~(\ref{parabolickinetichamiltonian}) can be written in terms of momenta:

                   \ba
\label{parabolickinetichamiltonianofp}
h_{K}(p)&=&\frac{p^2}{2}+p+\frac{1}{4}\,\,\,\,if\,\,\,\,p\geq -1/2\,\,\,\,\, (branch\,\,B_{+})\nonumber\\
&=&-p^2/2\,\,\,if\,\,\,\,\,1/2\geq p\geq -1/2 \,\,\,\,\,\,\,(branch\,\,B_{0})\nonumber\\
&=&\frac{p^2}{2}-p+\frac{1}{4}\,\,\,\,\,if\,1/2\geq p\,\,\,\, (branch\,\,B_{-})
\ea

		            \begin{figure}[t]
\vskip 0.3cm
                        \includegraphics[width=0.47\textwidth]{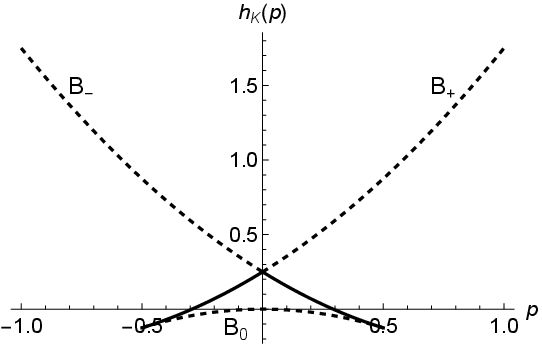}
                        \caption{\label{fig5}The piecewise parabolic kinetic part of the Hamiltonian $h_{K}(p)$ of Eq.~(\ref{parabolickinetichamiltonianofp}) as a function of momentum. The continuous lines show the branches that have the lower action.}
                        \end{figure}

and is shown in Figure~\ref{fig5}. We observe that it is multivalued. If, however, we impose the requirement that the branches followed by the system be those with the smaller value of the action, then the Hamiltonian becomes single-valued, as we shall show presently. In other words, at every intersection of two branches we demand that the system take the path where the action has the smaller value, for a small time interval around this intersection.

Indeed, let us examine each branch around an intersection point $p(t_{0})=p_0$ and $x(t_{0})=x_0$.

For the branch $B_{+}$ the solutions are

\ba
p(t)&=&-1+(1+p_{0})\cos(t-t_{0})-x_{0}\sin(t-t_0)\nonumber\\
x(t)&=&x_{0}\cos(t-t_0)+(1+p_0)\sin(t-t_0),
\ea
the corresponding action for the time interval $(t_0,t_0+\Delta t)$ being

\be
\Sigma_{+}\approx (2p_{0}^{2}-2x_{0}^{2}-1)\Delta t/4-(1+2p_{0})x_{0}\Delta t^{2}/2.
\ee

For the branch $B_{-}$ the solutions are

\ba
p(t)&=&1+(-1+p_{0})\cos(t-t_{0})-x_{0}\sin(t-t_0)\nonumber\\
x(t)&=&x_{0}\cos(t-t_0)+(-1+p_0)\sin(t-t_0),
\ea
the corresponding action for the time interval $(t_0,t_0+\Delta t)$ being

\be
\Sigma_{-}\approx (2p_{0}^{2}-2x_{0}^{2}-1)\Delta t/4+(1-2p_{0})x_{0}\Delta t^{2}/2.
\ee

For the branch $B_{0}$ the solutions are

\ba
p(t)&=&p_{0}\cosh(t-t_{0})-x_{0}\sinh(t-t_0)\nonumber\\
x(t)&=&x_{0}\cosh(t-t_0)-p_0\sinh(t-t_0),
\ea
the corresponding action for the time interval $(t_0,t_0+\Delta t)$ being

\be
\Sigma_{0}\approx (-p_{0}^{2}-x_{0}^{2})\Delta t/2+p_{0}x_{0}\Delta t^{2}.
\ee

Consequently, at the central intersection ($p_{0}=0)$ the difference in the actions is $\Sigma_{-}-\Sigma_{+}\approx x_{0}\Delta t^2$. Therefore the system will follow the path $B_{+}$ if $x_{0}>0$ and the path $B_{-}$ if $x_{0}<0$.

Similarly, at the intersection on the left ($p_{0}=-1/2$) the difference in the actions is $\Sigma_{0}-\Sigma_{+}\approx -x_{0}\Delta t^2/2$. Therefore the system will follow the path $B_{0}$ if $x_{0}>0$ and the path $B_{+}$ if $x_{0}<0$.

Finally, at the intersection on the right ($p_{0}=1/2$) the difference in the actions is $\Sigma_{0}-\Sigma_{-}\approx x_{0}\Delta t^2/2$. Therefore the system will follow the path $B_{-}$ if $x_{0}>0$ and the path $B_{0}$ if $x_{0}<0$.

Hence, the multivaluedness of the Hamiltonian as a function of momentum is removed and the equations of Hamilton can be used to find the solutions at all times.

Let us now examine the solutions of the system if the initial conditions are $x(0)=0$ and $v(0)=\epsilon+1/2$. This point lies on the $B_{+}$ branch, hence $p(0)=\epsilon-1/2$. The solution of the Hamilton equations for this branch are then

\ba
p(t)&=&-1+(\frac{1}{2}+\epsilon)\cos t\nonumber\\
x(t)&=&(\frac{1}{2}+\epsilon)\sin t,
\ea

{\bf (i) Case $\bf \epsilon<1/2$}

Let us examine first the case $\epsilon<1/2$. Then the first intersection the momentum will meet as it decreases is the one on the left, at $p=-1/2$. It will get there at time $t_{1}$, given by Eq.~(\ref{cost1}). At that point $\dot p(t_{1})=-x(t_{1})=-\sqrt{\epsilon^2+\epsilon}=-A$. Hence the momentum would want to keep decreasing. However, both the branches $B_{+}$ and $B_{0}$ that intersect there terminate at $p=-1/2$. The only possibility for the momentum is then to go to the branch $B_{-}$. Conservation of the energy $E=(\epsilon^2+\epsilon)/2-1/8$ forces it, in fact, to jump to the value $p=1/2$, while keeping the value of the position equal to $A$. As we just showed, the system will have then to follow the path with the smaller action, i.e. $B_{-}$.

The initial conditions for this branch will thus be $p(t_{1})=1/2$, $x(t_{1})=A$. The corresponding solutions are then

\ba
\label{eqatpisonmiso}
p(t)&=&1-(\epsilon+1/2)\cos(t-2t_{1})\nonumber\\
x(t)&=&-(\epsilon+1/2)\sin(t-2t_{1}).
\ea

Hence $p(3t_{1})=1/2$,  and $x(3t_{1})=-A$. Between the times $t=t_{1}$ and $t=3t_{1}$ the momentum is minimised at $t=2t_{1}$: $p(2t_{1})=1/2-\epsilon>0$. Hence it never reaches the central intersection. Having returned to the intersection on the right at $t=3t_{1}$, the momentum would like to keep increasing there, since $\dot p(3t_{1})=A>0$.  However this is impossible, since both the branch $B_{-}$ and the branch $B_{0}$ terminate at $p=1/2$. The only choice it has that will conserve energy is to jump to the intersection on the left, to the value $p=-1/2$, while the position still maintains the value $x(3t_{1})=-A$. As we determined before, the system will prefer to take the path of the smallest available action there, which is the branch $B_{+}$, since the position at $t=3t_{1}$ is negative.

We thus have to solve the equations of motion for the branch $B_{+}$ under the initial conditions $p(3t_{1})=-1/2$, $x(3t_{1})=-A$. We obtain

\ba
p(t)&=&-1+(\epsilon+1/2)\cos(t-4t_{1})\nonumber\\
x(t)&=&(\epsilon+1/2)\sin(t-4t_{1}).
\ea

\begin{figure}[t]
\vskip 0.3cm
                        \includegraphics[width=0.47\textwidth]{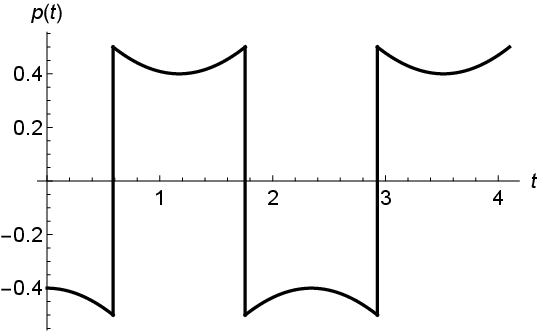}
                        \caption{\label{fig6} The periodic solution for the momentum $p(t)$ for $\epsilon=0.1$.}
                        \end{figure}

We see that $p(4t_{1})=\epsilon-1/2$, $p(5t_{1})=-1/2$ and $x(5t_{1})=A$. Thus the system returns at $t=5t_{1}$ to the intersection on the left, with a negative $\dot p$, wanting to keep decreasing below $-1/2$. However, since both $B_{+}$ and $B_{0}$ terminate at $p=-1/2$, the system will have to jump to $p=1/2$ in order to conserve the energy: $p(5t_{1})=1/2$, $x(5t_{1})=A$. Since the position is positive at that intersection, the path that has to be followed is $B_{-}$. We can solve thus Hamilton's equations for this branch, obtaining the solutions

\ba
p(t)&=&1-(\epsilon+1/2)\cos(t-6t_{1})\nonumber\\
x(t)&=&-(\epsilon+1/2)\sin(t-6t_{1}).
\ea

and so on. We obtain thus the momentum as a function of time shown in Figure~\ref{fig6}, for $\epsilon=0.1$. We note that the momentum jumps whenever its absolute value is 1/2. We have a continuous vacillation of $p$ between the values $1/2$ and $-1/2$. It lives on the parts of the branches shown in Figure~\ref{fig5} by continuous lines. The position is, of course, again the one shown in Figure~\ref{fig3}.

{\bf (ii) Case $\bf \epsilon>1/2$}

Let us again examine the solutions of the system if the initial conditions are $x(0)=0$ and $v(0)=\epsilon+1/2$. This point lies on the $B_{+}$ branch, hence $p(0)=\epsilon-1/2>0$. As the momentum decreases along the branch $B_{+}$ it will reach the central intersection point at time $t=\omega_{0}=\cos^{-1}(2/(1+2\epsilon))$, with $p(\omega_{0})=0$ and $x(\omega_{0})=C=\sqrt{\epsilon^2+\epsilon-3/4}$. As seen before, the favored path there is $B_{+}$, since the corresponding position is positive. So the system will keep moving along $B_{+}$ till it reaches the terminal $p=-1/2$ at time $t_{1}$, with $x(t_{1})=A$ and a negative $\dot p$. It would want to keep decreasing, but this is impossible, since both branches $B_{0}$ and $B_{+}$ end there. So the momentum jumps to the value $p=1/2$ on the intersection on the right, in order to conserve energy.

The corresponding initial values $p(t_{1})=1/2$ and $x(t_{1})=A$ indicate that the branch $B_{-}$ has to be followed now. The solutions are given by Eq.~(\ref{eqatpisonmiso}). Now, however, the momentum can reach the central intersection $p=0$ at time $t=2t_{1}-\omega_{0}$, at the position $x=C$. Since the position is positive, the system will have to switch branches at that moment in time and move to the branch $B_{+}$.

The solution of Hamilton's equations gives the solutions along the branch $B_{+}$:

\ba
p(t)&=&-1+(\epsilon+1/2)\cos(t-2t_{1}+2\omega_{0})\nonumber\\
x(t)&=&(\epsilon+1/2)\sin(t-2t_{1}+2\omega_{0}).
\ea

The momentum reaches the end of the branch at $p=-1/2$, $x=A$ and $t=3t_{1}-2\omega_{0}$, with a negative $\dot p$. The momentum would like to keep decreasing, but since we are at the end of the branches $B_{0}$ and $B_{+}$ it has to jump to the intersection on the right.

The new initial conditions $p(3t_{1}-2\omega_{0})=1/2$ and $x(3t_{1}-2\omega_{0})=A$ indicate that the branch with the lower action is now branch $B_{-}$. The corresponding solutions along this branch are

\ba
p(t)&=&1-(\epsilon+1/2)\cos(t-4t_{1}+2\omega_{0})\nonumber\\
x(t)&=&-(\epsilon+1/2)\sin(t-4t_{1}+2\omega_{0}).
\ea

\begin{figure}[t]
\vskip 0.3cm
                        \includegraphics[width=0.47\textwidth]{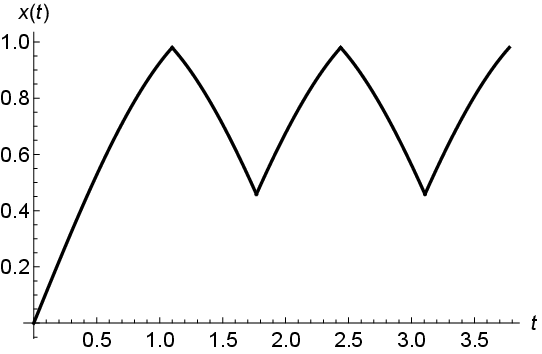}
                        \caption{\label{fig7} The periodic solution for the position $x(t)$ for $\epsilon=0.6$.}
                        \end{figure}

\begin{figure}[t]
\vskip 0.3cm
                        \includegraphics[width=0.47\textwidth]{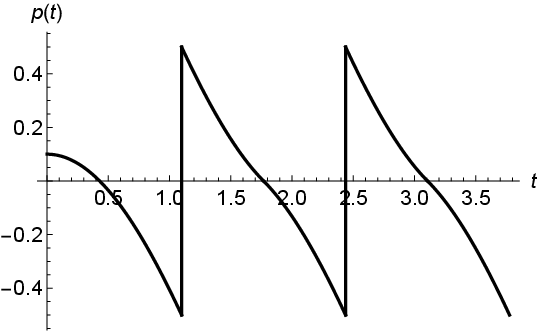}
                        \caption{\label{fig8} The periodic solution for the momentum $p(t)$ for $\epsilon=0.6$.}
                        \end{figure}

The system will keep moving along branches $B_{-}$ and $B_{+}$, due to the requirement that the action have always the lower available value. Thus we obtain the functions of time shown in Figure~\ref{fig7} and Figure~\ref{fig8}. These solutions have again periodic discontinuities in the momentum. They arose from the Hamilton equations of the single-valued Hamiltonian of the continuous branches shown in Figure~\ref{fig5}.

{\bf IV. Another piecewise parabolic Hamiltonian}

\begin{figure}[t]
\vskip 0.3cm
 \includegraphics[width=0.47\textwidth]{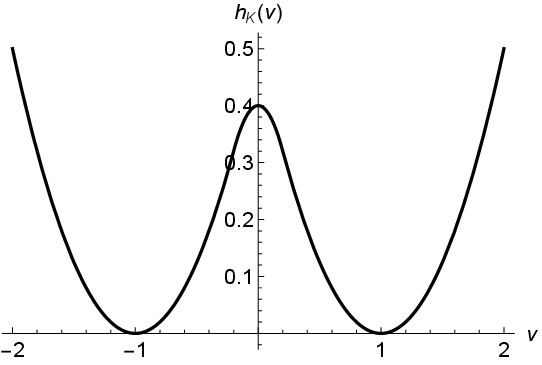}
\caption{\label{fig9}The piecewise parabolic kinetic part $h_{K}$  of Eq.~(\ref{parabolicHamiltoniankineticpart}) as a function of the speed for k=0.2.}
\end{figure}

Another piecewise defined model, which is also exactly solvable, involves another piecewise parabolic Hamiltonian. Its kinetic part is given by:

\ba
\label{parabolicHamiltoniankineticpart}
h_{K}(v)&=&(v-1)^2/2\,\,\,\,\,\,\,\,\,if\,\,\,\,v\geq k\nonumber\\
&=&\frac{(1-k)(k-v^2)}{2k}\,\,\,\,\,\,\,\,\,\,if\,\,\,\,\,k\geq v\geq -k\nonumber\\
&=&(v+1)^2/2\,\,\,\,\,if\,\,\,\,\,-k\geq v
\ea

where $k$ is small and positive. It is continuous in value and slope everywhere and it is shown for $k=0.2$ in Figure~\ref{fig9}. Since $\partial h_{K}/\partial v=0$ at $v=\pm 1$, we have an infinite acceleration there, as seen before. 
		
\begin{figure}[t]
\vskip 0.3cm
\includegraphics[width=0.47\textwidth]{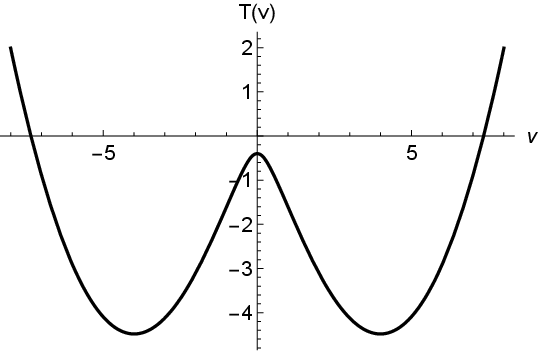}
\caption{\label{fig10} The piecewise parabolic kinetic part of Eq.~(\ref{parabolickineticwithlogs}) as a function of the speed for k=0.2.}                        \end{figure}

		The corresponding kinetic energy is shown in Figure~\ref{fig10}, it is continuous everywhere in both value and slope, has mirror symmetry and is equal to:
		
		\ba
\label{parabolickineticwithlogs}
T(v)&=&(v^2-1)/2+v\log(k/v)\,\,\,\,\,\,\,\,\,if\,\,\,\,v\geq k\nonumber\\
&=&(k-1)(k+v^2)/(2k)\,\,\,\,\,\,\,\,\,\,if\,\,\,\,\,k\geq v\geq -k\nonumber\\
&=&(v^2-1)/2+v\log(-v/k)\,\,if\,-k\geq v.
\ea

We note that $\partial^{2} T/\partial v^{2}=0$ at $v=\pm 1$, consequently we expect the acceleration to be infinite and the speed to be discontinuous there. We shall assume, without loss of generality, the initial conditions $x(0)=0$, $\dot x(0)=1+\epsilon$, with $\epsilon>0$. The corresponding energy is $\epsilon^{2}/2$. Hence $(\dot x(t)-1)^{2}/2+x(t)^{2}/2=\epsilon^{2}/2$, or equivalently
\be
\label{kladoi}
\dot x(t)=1\pm\sqrt{\epsilon^{2}-x(t)^{2}}.
\ee

This equation demonstrates that $|x(t)|\leq\epsilon$. We note that only the upper sign is consistent with the initial condition. The solution of this differential equation, with the initial condition $x(0)=0$, gives the time $t$ as an odd function $f(x)$ of the position:

\ba
\label{timefirst}
&&f(x)=\frac{\tan^{-1}(x/\sqrt{1-\epsilon^2})}{\sqrt{1-\epsilon^2}}+\tan^{-1}(x/\sqrt{\epsilon^2-x^2})\nonumber\\
&&-\frac{\tan^{-1}\left(\frac{x}{\sqrt{1 - \epsilon^2} \sqrt{\epsilon^2 - x^2}}\right)}{\sqrt{1 - \epsilon^2}}
\ea

The position starts at $x(0)=0$ and reaches its upper bound $\epsilon$ at time

\be
\label{t1}
t_{1}=f(\epsilon)=\frac{\tan^{-1}(\epsilon/\sqrt{1-\epsilon^2})}{\sqrt{1-\epsilon^2}}+\pi/2-\frac{\pi/2}{\sqrt{1 - \epsilon^2}}.
\ee

At that time the speed acquires the value 1 and the acceleration becomes infinite. Thus the speed has to make a jump to a value other than 1. Since the potential energy at that time has become equal to the total energy $\epsilon^{2}/2$, the kinetic part of the Hamiltonian has to be zero there. Hence the value of the speed after the jump can only be -1.

Then the equation of energy conservation after the time $t_{1}$ is $(\dot x(t)+1)^2/2+x(t)^2/2=\epsilon^2/2$. This leads to the possibilities $\dot x(t)=-1\pm\sqrt{\epsilon^2-x(t)^2}$. Which of the two will the system choose afterwards though? In order to find that, we shall assume that we end up with $\dot x(t)=-1+\nu\sqrt{\epsilon^2-x(t)^2}$, where $\nu^{2}=1$. Furthermore, the position is $\epsilon$ at time $t_{1}$, the speed being -1 then. The solution of this differential equation near $t_{1}$ is:

\be
\label{approximateposition}
x(t)\approx\epsilon-(t-t_{1})+\frac{\nu\sqrt{8\epsilon}}{3}(t-t_{1})^{3/2}.
\ee

The corresponding action for the interval $(t_{1},t_{1}+\Delta t)$ is then approximately

\be
\label{actiontochoose}
(-\epsilon^{2}/2+\log k)\Delta t-(\nu\sqrt{8\epsilon}/3)(\log k)\Delta t^{3/2}.
\ee
										
		In order to have the smallest available action, the choice $\nu=-1$ has to be made, since $k$ is small. In other words, the speed will go below -1 after the time $t_{1}$, with $\dot x(t)=-1-\sqrt{\epsilon^2-x(t)^2}$. Given the initial conditions $x(t_{1})=\epsilon$ and $\dot x(t_{1})=-1$, we can solve this particular differential equation to get the time as a function $g(x)$ of the position for $t\geq t_{1}$:

					\be
					\label{timesecond}
					g(x)=2t_{1}-f(x)
					\ee
									
									We see that $g(-\epsilon)=3t_{1}$, because $f(-\epsilon)=-t_{1}$. The position reaches therefore its lower bound $-\epsilon$ at time $3t_{1}$, with speed $-1$ and infinite acceleration. At that time the speed needs to jump once again. Since the potential energy has become once again equal to the total energy $\epsilon^{2}/2$, the kinetic energy has to be zero there. Hence the value of the speed after the jump can only be $+1$. Thus the equation of energy conservation after the time $3t_{1}$ is $(\dot x(t)-1)^2/2+x(t)^2/2=\epsilon^2/2$. This leads to the possibilities $\dot x(t)=1\pm\sqrt{\epsilon^2-x(t)^2}$.
										
In order to find which of the two will the system choose afterwards, we shall assume that $\dot x(t)=1+\nu\sqrt{\epsilon^2-x(t)^2}$, where $\nu^{2}=1$. Furthermore, the position is $-\epsilon$ at time $3t_{1}$, the speed being $+1$ then. The solution of this differential equation near $3t_{1}$ is:

\be
\label{approximateposition2}
x(t)\approx -\epsilon+(t-3t_{1})+\frac{\nu\sqrt{8\epsilon}}{3}(t-3t_{1})^{3/2}.
\ee

The corresponding action for the interval $(3t_{1},3t_{1}+\Delta t)$ is then approximately

\be
\label{actiontochoose2}
(-\epsilon^{2}/2+\log k)\Delta t+(\nu\sqrt{8\epsilon}/3)(\log k)\Delta t^{3/2}.
\ee
										
		In order to have the smallest available action, the choice $\nu=1$ has to be made, since $k$ is small. In other words, the speed will go above $+1$ after the time $3t_{1}$, with $\dot x(t)=1+\sqrt{\epsilon^2-x(t)^2}$. Given the initial conditions $x(3t_{1})=-\epsilon$ and $\dot x(3t_{1})=1$, we can solve the differential equation to get the time as a function $j(x)$ of the position for $t\geq 3t_{1}$:

					\be
					\label{timesecond3}
					j(x)=f(x)+4t_{1}
					\ee
									
\begin{figure}[t]
\vskip 0.3cm
                        \includegraphics[width=0.47\textwidth]{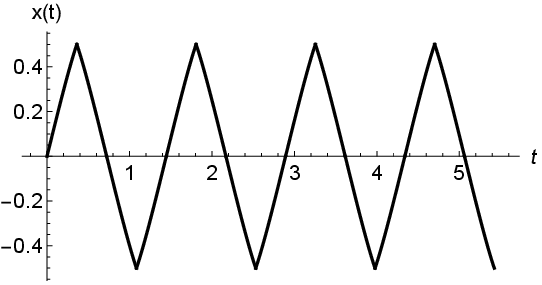}
                        \caption{\label{fig11}The position as a function of time for the piecewise parabolic Hamiltonian  of Eq.~(\ref{parabolicHamiltoniankineticpart}), for $\epsilon=0.5$.}
                        \end{figure}

\begin{figure}[t]
\vskip 0.3cm
                        \includegraphics[width=0.47\textwidth]{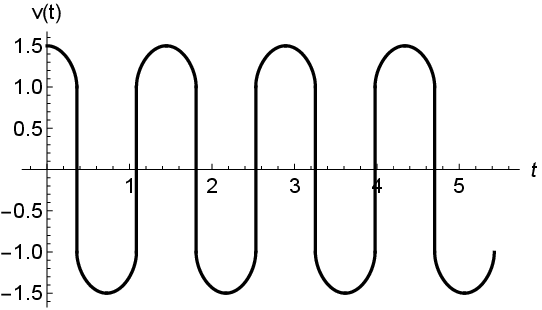}
                        \caption{\label{fig12}The speed as a function of time for the piecewise parabolic Hamiltonian  of Eq.~(\ref{parabolicHamiltoniankineticpart}), for $\epsilon=0.5$.}
                        \end{figure}

									We see that $j(-\epsilon)=3t_{1}$, because $f(-\epsilon)=-t_{1}$. The position reaches afterwards its upper bound $\epsilon$ at time $5t_{1}$, with speed $+1$ and infinite acceleration. At that time the speed jumps once again, and so on. The periodic solutions for the position and the speed are shown in Figure~\ref{fig11} and in Figure~\ref{fig12}, when $\epsilon=0.5$.

Thus the time as a function of the position is given by
										
\ba
\label{finaleq}
&&t=(-1)^{n}f(x)+2nt_{1}\nonumber\\
&&if\,\,\,(2n-1)t_{1}\leq t\leq(2n+1)t_{1}.
\ea

{\bf V. A quartic kinetic energy}
										
Let us examine now the quartic kinetic energy of Ref.\cite{Wilczek}:

\be
\label{quarticT}
T(v)=\frac{v^4}{4}-\frac{v^2}{2}.
\ee

The corresponding Hamiltonian is:

\be
\label{quarticH}
H=\frac{3v(t)^4}{4}-\frac{v(t)^2}{2}+x(t)^2/2.
\ee

Its kinetic part is

\be
\label{quarticHkinetic}
h_{K}(v)=\frac{3v(t)^4}{4}-\frac{v(t)^2}{2}.
\ee

The kinetic energy has $\partial^2 T/\partial v^2=0$ at $v=\pm 1/\sqrt{3}$, hence the acceleration is infinite and the speed will be discontinuous there. Furthermore, at these points the kinetic part of the Hamiltonian $h_{K}$ is minimised and takes the value $-1/12$.

The corresponding equation of motion is

\be
\label{equationquartic}
(3\dot x(t)^2-1)\ddot x(t)+x(t)=0.
\ee

Without loss of generality, we adopt the initial conditions $x(0)=0$ and $\dot x(0)=v_{0}$, with $v_{0}>1/\sqrt{3}$. Then the energy is

\be
\label{energyquartic}
E_{0}=\frac{3v_{0}^4}{4}-\frac{v_{0}^2}{2}.
\ee

\begin{figure}[t]
\vskip 0.3cm
                        \includegraphics[width=0.47\textwidth]{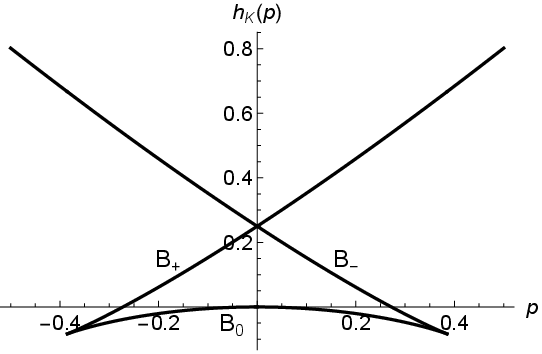}
                        \caption{\label{fig13}The kinetic part $h_{K}(p)$ of the quartic Hamiltonian  of Eq.~(\ref{quarticH}) as a function of momentum. }
                        \end{figure}

Whenever the speed is $\pm 1/\sqrt{3}$ the absolute value of the corresponding position is $b_{0}=(3v_{0}^{2}-1)/\sqrt{6}$, while the corresponding momentum is $\mp 2/(3\sqrt{3})$. We note that due to the nonlinearity of the momentum $p=v^3-v$, the Hamiltonian cannot be expressed easily in terms of the momentum. In fact, $H(p)$ is multi-valued, as shown in Figure~\ref{fig13}. Furthermore, the values $1$ and $-1$ of the speed correspond to $p=0$ and $h_{K}=1/4$.

Eq.~(\ref{equationquartic}) is transformed into the equation of energy conservation when multiplied by $\dot x(t)$ and integrated with respect to time. It is convenient to introduce the variable $c(t)$, where $x(t)=b\sin c(t)$. We can use energy conservation in order to find $dt/dc$ as a function of $c(t)$. We can then find analytically the time as a function of this new variable $c(t)$. The speed and the position can be related to time through the parameter $c(t)$. The corresponding parametric relations will involve elliptic integrals, making thus the solution look rather inscrutable.

However, we can use a different method for examining the quartic hamiltonian: we shall replace it by a piecewise parabolic kinetic energy, ensuring that the main characteristics of the quartic kinetic energy are incorporated into the parabolic one. We shall find that the resulting simple solutions of the new kinetic energy coincide quite well with the exact solutions of the quartic one, which involve elliptic integrals.

The deployment of exactly solvable piecewise models to substitute or approximate transcendental and non-smooth systems has a long standing precedent in mathematical physics\cite{Znojil}, but only for the potential energy, not the kinetic one. Extending this approach, we construct an emulated piecewise parabolic kinetic energy that mirrors the critical structural characteristics of the original quartic framework while maintaining analytical tractability. Actually, this method is quite effective in more general settings as well and it can be used in numerical calculations, where the trajectories of a Lagrangian system can be approximated by replacing the original Lagrangian by an approximate Lagrangian that can be solved exactly\cite{Gonzalez}.

For example, let us look at the relativistic harmonic oscillator, where the dimensionless Lagrangian is $-\sqrt{1-v^2}-x^2/2$. Let us adopt, for the sake of being specific, the initial conditions $x(0)=0$ and $v(0)=v_{0}$, the corresponding energy and maximal position being $E=1/\sqrt{1-v_{0}^2}$ and $\sqrt{2E-2}$, respectively. The exact solution involves complicated elliptic integrals. However, we can emulate the kinetic energy by a parabolic function having the same structure. This is the function $-1+(1-\sqrt{1-v_{0}^2})(v/v_{0})^2$, which has the same value and slope at $v=0$ and the same value at $v=v_{0}$. The solution of the parabolic Lagrangian is $x(t)=w\sin(tv_{0}/w)$, where $w=\sqrt{2-2/E}$. Consequently, the period of the oscillation is $(4\pi/v_{0})\sqrt{(E-1)/(2E)}$. This is a simple and quite accurate approximation to the exact period. For example, if $v_{0}=1/2$ the exact period is 6.64, while the emulated one is 6.50.

Let us adopt then an emulated piecewise parabolic kinetic part of the Hamiltonian, in the place of the quartic one:

\ba
\label{parabolickineticforquartic}
&&h_{+}(p)=a_{1}p^2+b_{1}p+c_{1}\,\,\,if\,\,\,\,p\geq -2/(3\sqrt{3})\nonumber\\
&&h_{0}(p)=a_{2}p^2+c_{2}\,\,\,\,if\,\,\,\,\,2/(3\sqrt{3})\geq p\geq -2/(3\sqrt{3})\nonumber\\
&&h_{-}(p)=a_{1}p^2-b_{1}p+c_{1}\,\,\,if\,\,\,\,p\leq 2/(3\sqrt{3})
\ea

These kinetic parts for the branches $B_{+}$, $B_{0}$ and $B_{-}$ reespectively have to emulate the kinetic part of the quartic Hamiltonian that is shown in Figure~\ref{fig13}. So, when $p=-2/(3\sqrt{3})$ the value of $h_{+}(p)$ must be $-1/12$ and the corresponding speed $h_{+}^{\prime}(p)$ must equal $1/\sqrt{3}$, while $h_{+}(0)=1/4$. These three conditions yield the parameters $a_{1}=3/4$, $b_{1}=2/\sqrt{3}$ and $c_{1}=1/4$ for the branch $B_{+}$.

Similarly for the branch $B_{0}$ we must have $h_{0}(-2/(2\sqrt{3}))=-1/12$ and $h_{0}^{\prime}(-2/(2\sqrt{3}))=1/\sqrt{3}$. Hence we must have $a_{2}=-3/4$, $c_{2}=1/36$.

\begin{figure}[t]
\vskip 0.3cm
                        \includegraphics[width=0.47\textwidth]{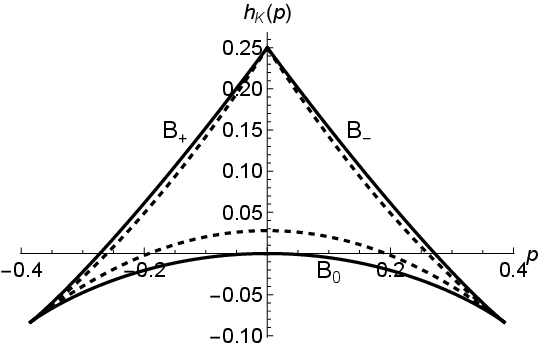}
                        \caption{\label{fig14}The kinetic part $h_{K}(p)$ of the quartic Hamiltonian  (continuous lines) and of its piecewise parabolic simulacrum (dashed lines) as a function of momentum.}
                        \end{figure}

Thus we obtain the hamiltonian simulacrum shown with dashed lines in Figure~\ref{fig14}:

\ba
\label{parabolickineticforquarticagain}
h_{K}&=&\frac{3}{4}p^2+\frac{2}{\sqrt{3}}p+\frac{1}{4}\,\,\,if\,\,\,\,p\geq -2/(3\sqrt{3})\nonumber\\
&=&-\frac{3}{4}p^2+\frac{1}{36}\,\,\,\,if\,\,\,\,\,2/(3\sqrt{3})\geq p\geq -2/(3\sqrt{3})\nonumber\\
&=&\frac{3}{4}p^2-\frac{2}{\sqrt{3}}p+\frac{1}{4}\,\,\,if\,\,\,\,p\leq 2/(3\sqrt{3})
\ea

                        The solutions of the Hamilton equations of motion for the branch $B_{+}$, given the initial conditions $p(t_{0})=p_{0}$ and $x(t_{0})=x_{0}$, are:

\ba
\label{B+sols}
p(t)&=&\frac{4}{3\sqrt{3}}(\cos\phi-1)+p_{0}\cos\phi-x_{0}\sqrt{\frac{2}{3}}\sin\phi\nonumber\\
x(t)&=&x_{0}\cos\phi+\frac{(4+3\sqrt{3}p_{0})\sin\phi}{3\sqrt{2}},
\ea

where $\phi=\sqrt{3/2}(t-t_{0})$.
The corresponding action for the interval $(t_{0},t_{0}+\Delta t)$, where $\Delta t$ is positive, is

\be
\label{B+action}
s_{+}\approx\frac{1}{4}(-1+3p_{0}^{2}-2x_{0}^{2})\Delta t-(\frac{x_{0}}{\sqrt{3}}+\frac{3}{2}p_{0}x_{0})\Delta t^2.
\ee

                        The solutions of the Hamilton equations of motion for the branch $B_{-}$, given the initial conditions $p(t_{0})=p_{0}$ and $x(t_{0})=x_{0}$, are:

\ba
\label{B-sols}
p(t)&=&\frac{4}{3\sqrt{3}}(1-\cos\phi)+p_{0}\cos\phi-x_{0}\sqrt{\frac{2}{3}}\sin\phi\nonumber\\
x(t)&=&x_{0}\cos\phi+\frac{(3\sqrt{3}p_{0}-4)\sin\phi}{3\sqrt{2}}.
\ea

The corresponding action for the interval $(t_{0},t_{0}+\Delta t)$, where $\Delta t$ is positive, is

\be
\label{B-action}
s_{-}\approx\frac{1}{4}(-1+3p_{0}^{2}-2x_{0}^{2})\Delta t+(\frac{x_{0}}{\sqrt{3}}-\frac{3}{2}p_{0}x_{0})\Delta t^2.
\ee

The solutions of the Hamilton equations of motion for the branch $B_{0}$, given the initial conditions $p(t_{0})=p_{0}$ and $x(t_{0})=x_{0}$, are:

\ba
\label{B0sols}
p(t)&=&p_{0}\cosh\phi-x_{0}\sqrt{\frac{2}{3}}\sinh\phi\nonumber\\
x(t)&=&x_{0}\cosh\phi-\sqrt{\frac{3}{2}}p_{0}\sinh\phi.
\ea

The corresponding action for the interval $(t_{0},t_{0}+\Delta t)$, where $\Delta t$ is positive, is

\be
\label{B0action}
s_{0}\approx\frac{1}{36}(-1-27p_{0}^{2}-18x_{0}^{2})\Delta t+\frac{3}{2}p_{0}x_{0}\Delta t^2.
\ee

We can use these solutions to track the evolution of the system when it starts out at $x(0)=0$ with $\dot x(0)=v_{0}$, where $1/\sqrt{3}<v_{0}<1$. It is initially on the branch $B_{+}$, so the corresponding solutions are those of Eq.~(\ref{B+sols}) for the values $t_{0}=0$, $x_{0}=0$ and $p_{0}=v_{0}^3-v_{0}$:

\ba
\label{B+sols1}
p_{1}(t)&=&-\frac{4}{3\sqrt{3}}+(\frac{4}{3\sqrt{3}}-v_{0}+v_{0}^3)\cos(t\sqrt{3/2})\nonumber\\
x_{1}(t)&=&\frac{4+3\sqrt{3}(-v_{0}+v_{0}^3)}{3\sqrt{2}}\sin(t\sqrt{3/2}).
\ea

The motion on the $B_{+}$ branch ends at time $t=\tau_{1}$, at the intersection on the left in Figure~\ref{fig14}, where the speed acquires the value $1/\sqrt{3}$ and the momentum becomes $-2/(3\sqrt{3})$:

\be
\label{tau1}
\tau_{1}=\sqrt{2/3}\sec^{-1}(2+3\sqrt{3}(v_{0}^3-v_{0})/2).
\ee

At that point the system is at the position $x(\tau_{1})=b$, where

\be
\label{bpros}
b=\sqrt{\frac{2}{3}-\frac{4v_{0}}{\sqrt{3}}+\frac{3v_{0}^2}{2}+\frac{4v_{0}^3}{\sqrt{3}}-3v_{0}^4+\frac{3}{2}v_{0}^6}.
\ee

\begin{figure}[t]
\vskip 0.3cm
                        \includegraphics[width=0.47\textwidth]{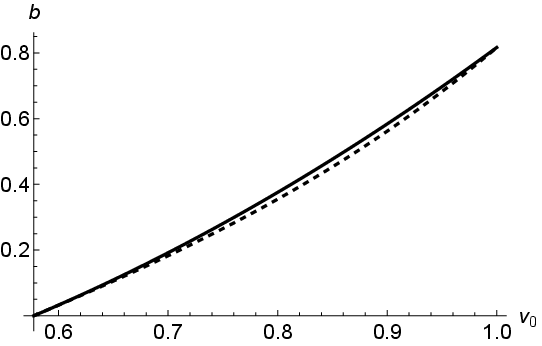}
                        \caption{\label{fig15}The maximum displacement from the equilibrium position $x=0$  for the quartic Hamiltonian as a function of the initial speed $v_{0}$. The continuous line corresponds to the exact value $b_{0}$, while the simulacrum of Eq.~(\ref{bpros}) is given by the dashed line.}
                        \end{figure}

This simulacrum of the maximum displacement of the system from the equilibrium position $x=0$ agrees very well with the exact value $b_{0}=(3v_{0}^{2}-1)/\sqrt{6}$ mentioned earlier, as we can see in Figure~\ref{fig15}.

                        After reaching the endpoint of the branch $B_{+}$ the system will jump to the intersection of branches $B_{0}$ and $B_{-}$ on the right, with momentum $2/(3\sqrt{3})$, speed $-1/\sqrt{3}$ and position $b$. This will be the starting point for the second part of the motion. The preferred branch for this part of the motion is the branch $B_{-}$, since the difference of the corresponding actions is $s_{0}-s_{-}=b\Delta t^{2}/\sqrt{3}$. Thus the system will move along the branch that has the smaller action between the two.
                        
                        The corresponding solutions for the $B_{-}$ branch are

\ba
\label{B-sols2}
p_{2}(t)&=&\frac{4}{3\sqrt{3}}-\frac{2}{3\sqrt{3}}\cos\phi_{2}-b\sqrt{\frac{2}{3}}\sin\phi_{2}\nonumber\\
x_{2}(t)&=&b\cos\phi_{2}-\frac{\sqrt{2}}{3}\sin\phi_{2},
\ea

where $\tau_{1}<t<3\tau_{1}$ and $\phi_{2}=\sqrt{3/2}(t-\tau_{1})$.

We note that $\dot p_{2}(2\tau_{1})=0$, $x_{2}(2\tau_{1})=0$, $p_{2}(3\tau_{1})=2/(3\sqrt{3})$, $x_{2}(3\tau_{1})=-b$ and $\dot x_{2}(3\tau_{1})=-1/\sqrt{3}$. Thus the system proceeds along the branch $B_{-}$ till it reaches a positive minimum momentum and then it returns to the intersection on the right of Figure~\ref{fig14} at the instant $3\tau_{1}$, provided $v_{0}<1$. Thus it never reaches the top intersection, the one between the branches $B_{+}$ and $B_{-}$. Note that here we assumed $v_{0}<1$, so that the intersection at the top is not reached.

If we had allowed $v_{0}>1$, the system would have reached the intersection at the top of Figure~\ref{fig14} (where $p=0$) before the time $2\tau_{1}$, at a positive value of the position. There it would have chosen the path to take according to the principle of the smaller available action. Since $s_{-}-s_{+}=2x_{0}\Delta t^{2}/\sqrt{3}$ there, the system would have switched branches at that point and would have proceeded down the branch $B_{+}$.

\begin{figure}[t]
\vskip 0.3cm
                        \includegraphics[width=0.47\textwidth]{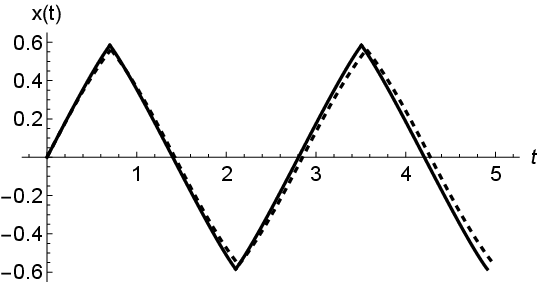}
                        \caption{\label{fig16}The solution for the position of the system as a function of time for the quartic Hamiltonian, for $v_{0}=0.9$. The continuous line corresponds to the exact solution, while its simulacrum is given by the dashed line.}
                        \end{figure}

\begin{figure}[t]
\vskip 0.3cm
                        \includegraphics[width=0.47\textwidth]{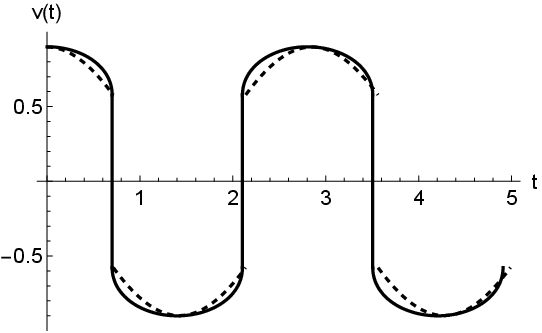}
                        \caption{\label{fig17}The solution for the speed of the system as a function of time for the quartic Hamiltonian, for $v_{0}=0.9$. The continuous line corresponds to the exact solution, while its simulacrum is given by the dashed line.}
                        \end{figure}

For the sake of concreteness, though, we have assumed that $v_{0}<1$, so the intersection at the top will never be reached. At the moment $3\tau_{1}$ the system has reached the endpoint of branch $B_{-}$ and therefore jumps to the intersection on the left of Figure~\ref{fig14}, preserving the energy. There starts the third part of the motion, with initial conditions $t_{0}=3\tau_{1}$, $x_{0}=-b$, $p_{0}=-2/(3\sqrt{3})$. Two branches are available there, $B_{+}$ and $B_{0}$, but the system will choose $B_{+}$, because $s_{0}-s_{+}=b\Delta t^2/\sqrt{3}$. The corresponding solutions along $B_{+}$ are given by Eq.~(\ref{B+sols}):

\ba
\label{B+sols3}
p_{3}(t)&=&-\frac{4}{3\sqrt{3}}+(\frac{2}{3\sqrt{3}})\cos\phi_{3}+b\sqrt{\frac{2}{3}}\sin\phi_{3}\nonumber\\
x_{3}(t)&=&\frac{\sqrt{2}}{3}\sin\phi_{3}-b\cos\phi_{3},
\ea

where $3\tau_{1}<t<5\tau_{1}$ and $\phi_{3}=\sqrt{3/2}(t-3\tau_{1})$. We note that $p_{3}(5\tau_{1})=-2/(3\sqrt{3})$, $x_{3}(5\tau_{1})=b$, $\dot x_{3}(5\tau_{1})=1/\sqrt{3}$. We can continue like this, obtaining at the end the periodic solutions for the position and the speed shown in  Figure~\ref{fig16} and Figure~\ref{fig17} with dashed lines. These coincide fairly well with the exact solutions that involve elliptic integrals (shown with continuous lines).

\begin{figure}[t]
\vskip 0.3cm
                        \includegraphics[width=0.47\textwidth]{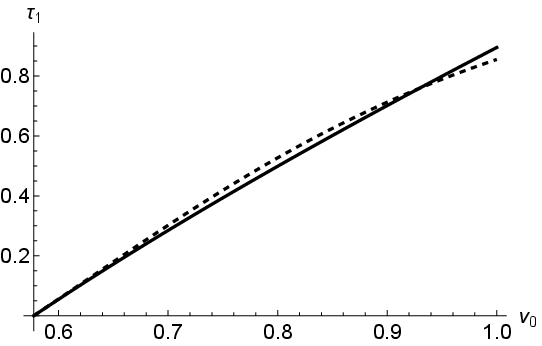}
                        \caption{\label{fig18}One quarter of the period of the solution  for the quartic Hamiltonian as a function of the initial speed $v_{0}$. The continuous line corresponds to the exact value, while the simulacrum of Eq.~(\ref{tau1}) is given by the dashed line.}
                        \end{figure}

The parameter $\tau_{1}$ is $1/4$ of the period of the solution. As we can see in Figure~\ref{fig18}, this coincides rather well with the exact solution that involves elliptic integrals.

{\bf VI. Conclusions}

Time crystals appear whenever the curvature of the kinetic part of the Lagrangian changes its sign somewhere. We have demonstrated the existence of such crystals in the case of a piecewise parabolic kinetic energy and a piecewise parabolic Hamiltonian, assuming in all cases a simple harmonic oscillator potential  for the position. The parabolic feature of the kinetic energy enables the equations of motion to be solved exactly. The solutions are periodic. Every time the speed jumps, the next path that has to be followed is the stable one and has the lower action. We have shown that, even in cases where the kinetic energy has very complicated solutions, we can examine all their features in detail and with sufficient accuracy if we replace their kinetic energies by parabolic simulacra. This replacement of the original equations of motion with emulated ones can be used quite generally, actually, leading to solvable equations of motion and quite useful insights into the details of the problem.

\end{document}